\documentclass[ijoc,nonblindrev]{informs3} 
\usepackage{graphicx}
\usepackage{float}
\usepackage[lofdepth,lotdepth]{subfig}
\usepackage{algorithm}
\usepackage[noend]{algpseudocode}
\usepackage{tikz}
\usetikzlibrary{shapes,arrows}
\usepackage{pifont}
\usepackage[round]{natbib}
\usepackage{amssymb}
\usepackage{lineno}
\usepackage{amsmath}
\usepackage{amsfonts}
\usepackage{booktabs}
\usepackage{textcomp}
\usepackage{hyperref}
\usepackage{multirow}
\usepackage{color}
\usepackage{natbib}
\usepackage[justification=centering]{caption}

 \bibpunct[, ]{(}{)}{,}{a}{}{,}%
 \def\bibfont{\small}%
\TheoremsNumberedThrough     

\EquationsNumberedThrough    

\begin{document}



\RUNTITLE{Identifying intercity freight trip ends of heavy trucks from GPS data}

\TITLE{Identifying intercity freight trip ends of heavy trucks from GPS data}

\ARTICLEAUTHORS{%
\AUTHOR{Yitao Yang}
\AFF{Key Laboratory of Integrated Transport Big Data Application Technology for Transport Industry, Beijing Jiaotong University, Beijing 100044, China, \EMAIL{yitao-yang@bjtu.edu.cn}}
\AUTHOR{Bin Jia\footnote{Corresponding author}}
\AFF{Institute of Transportation System Science and Engineering, Beijing Jiaotong University, Beijing 100044, China, \EMAIL{bjia@bjtu.edu.cn}}
\AUTHOR{Xiao-Yong Yan\footnote{Corresponding author}}
\AFF{Institute of Transportation System Science and Engineering, Beijing Jiaotong University, Beijing 100044, China, \EMAIL{yanxy@bjtu.edu.cn}}
\AUTHOR{Jiangtao Li }
\AFF{Institute of Transportation System Science and Engineering, Beijing Jiaotong University, Beijing 100044, China, \EMAIL{17125670@bjtu.edu.cn}}
\AUTHOR{Zhenzhen Yang}
\AFF{Institute of Transportation System Science and Engineering, Beijing Jiaotong University, Beijing 100044, China, \EMAIL{zzyang@bjtu.edu.cn}}
\AUTHOR{Ziyou Gao}
\AFF{Institute of Transportation System Science and Engineering, Beijing Jiaotong University, Beijing 100044, China, \EMAIL{zygao@bjtu.edu.cn}}
} 

\ABSTRACT{%
The intercity freight trips of heavy trucks are important data for transportation system planning and urban agglomeration management. In recent decades, the extraction of freight trips from GPS data has gradually become the main alternative to traditional surveys. Identifying the trip ends (origin and destination, OD) is the first task in trip extraction. In previous trip end identification methods, some key parameters, such as speed and time thresholds, have mostly been defined on the basis of empirical knowledge, which inevitably lacks universality. Here, we propose a data-driven trip end identification method. First, we define a speed threshold by analyzing the speed distribution of heavy trucks and identify all truck stops from raw GPS data. Second, we define minimum and maximum time thresholds by analyzing the distribution of the dwell times of heavy trucks at stop location and classify truck stops into three types based on these time thresholds. Third, we use highway network GIS data and freight-related points- of- interest (POIs) data to identify valid trip ends from among the three types of truck stops. In this step, we detect POI boundaries to determine whether a heavy truck is stopping at a freight-related location. We further analyze the spatiotemporal characteristics of intercity freight trips of heavy trucks and discuss their potential applications in practice.
}%


\maketitle

\section{Introduction}
In recent years, global production activities and the rapid development of e-commerce have accelerated the growth of intercity freight demand [1, 2]. According to the UN Department of Economic and Social Affairs, the world population is expected to increase by approximately 29\% by 2050 [3]. Intercity freight demand will be further stimulated in the future due to population growth, urbanization and industrialization. Heavy trucks are mainly used to transport goods between manufacturers and distribution centers in intercity freight [4, 5]. However, an increase in the number of heavy trucks on the road may lead to congestion at bottlenecks, reduce the traffic flow speed and increase traffic accidents [6-8]. Therefore, traffic departments need to effectively supervise heavy trucks to improve the operational efficiency of road networks. However, a major challenge at present is the lack of data on the intercity freight trips of heavy trucks, which hinders a deep understanding of freight transport systems [9].

Traditionally, data on the intercity freight trips of heavy trucks are obtained through manual surveys (investigations, work diaries, etc.) [10], which are costly and time consuming [11, 12]. In recent decades, GPS surveys have gradually become the main alternative to traditional surveys as an increasing number of freight companies have adopted GPS technology for fleet management [13, 14]. GPS data have been widely applied in freight modeling applications [14-18]. However, GPS data contain only information on geographic location and truck status and do not include trip descriptions [19]. Therefore, the first task in GPS-data-driven freight modeling is to extract freight trips by identifying the trip ends (loading stops, unloading stops and rest stops) of heavy trucks [12].

The process of identifying heavy truck trip ends consists of two main steps: (1) identifying truck stops from raw GPS data and (2) selecting valid trip ends from among the identified truck stops. For the first step, previous studies [13] have suggested that the instantaneous speed of a heavy truck and secondary data (engine on/off status, ignition information, etc.) can be used to identify truck stops. However, such secondary data are often difficult to obtain in practice. Instead, other previous studies [9, 12, 19-21] have used the average speed between successive GPS points to infer the motion status (moving or stationary) of a heavy truck, for which a corresponding speed threshold is a key predefined parameter. In this method, the motion status of a heavy truck is considered to be stationary if the calculated average speed between successive GPS points is lower than the defined speed threshold. Depending on the specific research background, various speed thresholds have been selected in previous studies, such as 14 km/h [22], 8 km/h [9], 5 km/h [21] and 1 km/h [12]. The main deficiency of these existing studies is that the speed thresholds have mostly been defined on the basis of practical experience. A more universal method of defining the speed threshold needs to be further studied.

For selecting valid trip ends from among identified truck stops, three main methods have been applied in existing studies: (1) the geographic information method [23-26], (2) the spatial clustering method [24, 27-29], and (3) the dwell time method [12, 14, 30, 31]. The geographic information method [26] mainly considers the spatial relationship between truck stops and points (or regions) of interest, such as freight enterprises. However, it is impossible to obtain all relevant geographic information in advance, and constructing large geographic databases is costly and time consuming. Therefore, the geographic information method is seldom used in practice.

The spatial clustering method [32-34] mainly considers the spatial characteristics of heavy truck trajectories. Truck stops that satisfy certain spatial density constraints are clustered into a group, which corresponds to a trip end. Recently, a variety of spatial clustering algorithms have been proposed, such as an improved K-means algorithm [35], DJ-Cluster [27], CB-SMoT [24], T-DBSCAN [29], DBSCAN-TE [36], and TAD [26]. Overall, the spatial clustering method can effectively identify the trip ends and other locations frequently visited by heavy trucks. However, this method has high requirements in terms of GPS data quality. Data loss, data drift and inconsistent sampling rates will affect the spatial distribution of the GPS points. Thus, continuous trajectories may be inaccurately clustered into multiple groups due to anomalously low densities [19, 37].

The dwell time method [12, 14, 20, 30, 31, 38] mainly considers the time characteristics of heavy truck trajectories. In this method, the time thresholds are key parameters that need to be predefined [12]. If a heavy truck remains at a certain location longer than a specific time threshold, then the corresponding truck stop is identified as a trip end. Generally, the loading/unloading times of heavy trucks are dozens of minutes or hours, while the wait time for a signal light is relatively short (mostly within a few minutes). In previous studies, short time thresholds of 3 min [14, 39] and 5 min [31] have been selected in accordance with the traffic signal cycle, refueling time and road traffic conditions. However, 3 or 5 min may be too short to complete one loading or unloading process for a heavy truck due to the particular characteristics of heavy truck freight activities [40]. Therefore, some studies have suggested that longer time thresholds, such as 10 min [41], 15 min [12, 20] or 20 min [30], should be selected. In addition, urban freight policy is an important factor to consider. For example, multiple segmental time thresholds are defined in accordance with the maximum rest time for a single freight trip in the European Union [42]. Overall, the dwell time method has low complexity and is simple to implement [19]. However, the main deficiency of previous studies is that the time thresholds have mostly been defined on the basis of practical experience. Moreover, few studies have addressed abnormal events such as long periods of traffic congestion. A truck stop may be inaccurately identified as a trip end if the heavy truck is stuck on the road longer than the selected time threshold.

In this paper, we propose a data-driven trip end identification method. Our dataset contains data from approximately 2.6 million heavy trucks in China. The proposed method of identifying heavy truck trip ends from GPS data consists of two steps. The first step is to identify truck stops from the GPS data, for which we define a speed threshold by analyzing the speed distribution of heavy trucks. The second step is to select valid trip ends from among the identified truck stops, for which we adopt the dwell time method. First, we define time thresholds by analyzing the distribution of the dwell times of heavy trucks at stop locations and classify truck stops into three types based on these time thresholds. Second, we analyze the spatial relationship between truck stops and freight-related points –of -interest (POIs) to detect the POI boundaries, which are used to determine whether a heavy truck is stopping at a certain location. Third, we use highway network GIS data and freight-related POI data to identify valid trip ends from among the three types of truck stops.

\section{Data description}

\subsection{Data source}
For this study, we obtained GPS data from the National Road Freight Supervision and Service Platform (https://www.gghypt.net/). This platform is mainly used by the government to monitor road violations (speeding, fatigue driving, etc.) of heavy trucks. According to the regulations of the Ministry of Transport of China, heavy trucks with a load exceeding 12 tons have been required to be equipped with vehicle positioning equipment since July 2014. The trajectory information of a heavy truck is uploaded to the above platform in real time. Our GPS dataset contains data from approximately 2.6 million heavy trucks, accounting for approximately 41\% of the total in China, and the number of records is greater than 20 billion (see Table 1 for examples). The time span of the GPS data is one week (from May 18 to May 24, 2018), and the average sampling rate is 30 s.

\begin{table}[htbp]
	\small
	\centering
	\caption{Examples of heavy truck GPS data}
	\begin{tabular}{ccccccc}
		\toprule
		Index  & ID & Longitude & Latitude & Speed & Timestamp & Direction \\
		\midrule
		1 & 60817be2749c77 & 119.786484 & 34.387562 & 0     & 2018-05-19 12:03:20 & 132       \\
		2 & 60817be2749c77 & 119.787315 & 34.388016 & 30    & 2018-05-19 12:03:50 & 169       \\
		3 & 60817be2749c77 & 119.788536 & 34.388783 & 25    & 2018-05-19 12:04:20 & 70        \\
		4 & 60817be2749c77 & 119.789902 & 34.38847  & 7     & 2018-05-19 12:04:50 & 206       \\
		5 & 60817be2749c77 & 119.789902 & 34.38847  & 0     & 2018-05-19 12:05:20 & 206       \\
		\bottomrule
	\end{tabular}%
	\label{tab:small_problem}%
\end{table}%

\subsection{Data preprocessing}
Raw GPS data contain considerable erroneous and redundant information [43]. For example, when a heavy truck is inside a city canyon, data jumps and data drift may occur as the GPS signal is reflected or obstructed by buildings. The GPS signal loss is particularly significant when a heavy truck enters a tunnel. The corresponding abnormal data need to be removed through data preprocessing. We consider three main types of abnormal data: (1) duplicate or missing data, (2) unreasonable data, and (3) data jumps. For duplicate or missing data, we directly remove the corresponding GPS records to ensure data authenticity and validity. For unreasonable data, such as GPS points outside the national border, we also remove the corresponding GPS records. For data jumps, we first calculate the average speed and acceleration among successive GPS points and then compare them against a defined maximum speed (120 km/h) and acceleration (5 m/s2). A data jump is considered to have occurred if the average speed or acceleration is larger than the respective maximum value, and the corresponding GPS records are removed.

One thing to note is that the time intervals between some pairs of successive GPS points may become too large as a result of data preprocessing and signal loss. It is difficult to determine the activities of heavy trucks in excessively long time intervals without available data, which will limit the application value of such data. Referring to previous studies [13, 20], we define a long time interval threshold of 1 h. If a freight trip extracted from heavy truck GPS data contains two successive records separated by a time interval of more than 1 h, that freight trip will be discarded.

\section{Methodology}
The proposed method of heavy truck freight trip end identification mainly consists of two key steps, as illustrated in Fig. 1. The first step is to identify truck stops from raw GPS data, and the second step is to select valid trip ends from among the identified truck stops.

\begin{figure}
	\centering
	\includegraphics[scale=0.5]{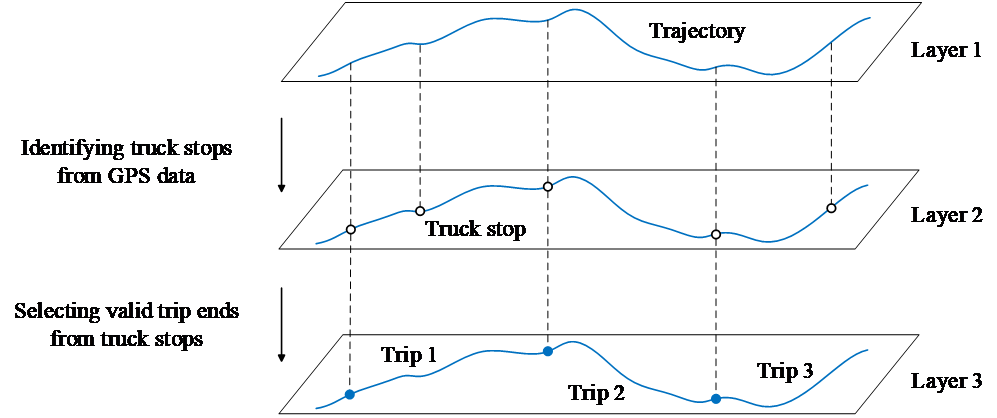}
	\captionsetup{justification=centering}
	\caption{Illustration of the proposed methodology. The first layer represents a trajectory after data preprocessing. The second layer represents the truck stops identified from the preprocessed trajectory. The third layer represents the valid trip ends selected from among all truck stops, which are used to divide the continuous trajectory into multiple trips. The white hollow dots and blue solid dots represent truck stops and trip ends, respectively. The dashed black lines represent the location correspondences between the layers.}
\end{figure}

\subsection{Identifying truck stops from GPS data}
We calculate the average speed between two successive GPS points to infer the motion status (moving or stationary) of the heavy truck at each GPS point. GPS point i is called a stopped GPS point if the heavy truck is stationary at time $t_i$. We can identify truck stops from successive stopped GPS points (as shown in Fig. 2). The average speed from GPS point i to GPS point i+1 (deno$t_{i+1}$ted by $v_{i+1}$) is calculated as
\begin{flushright}
	$v_{i+1}$=$d_{i,i+1}$/($t_{i+1}-t_{i}$)\qquad\qquad\qquad\qquad\qquad\qquad\quad(1)
\end{flushright} 
where $d_{i,i+1}$ denotes the geographic distance between GPS points i and i+1. The motion status of the heavy truck is stationary at time $t_{i+1}$ if $v_{i+1}{\leq}{\delta}_v$, where the speed threshold ${\delta}_v$ is a predefined parameter. Defining a suitable speed threshold is a difficult task due to the occurrence of data drift, which can cause the speed between two successive GPS points to take a random nonzero value even if the heavy truck is stationary. Moreover, the speed of a heavy truck on a congested road will be very low and may be close to the speed associated with data drift. Therefore, the motion characteristics of heavy trucks are important factors to consider when defining the speed threshold.

\begin{figure}
	\centering
	\includegraphics[scale=0.05]{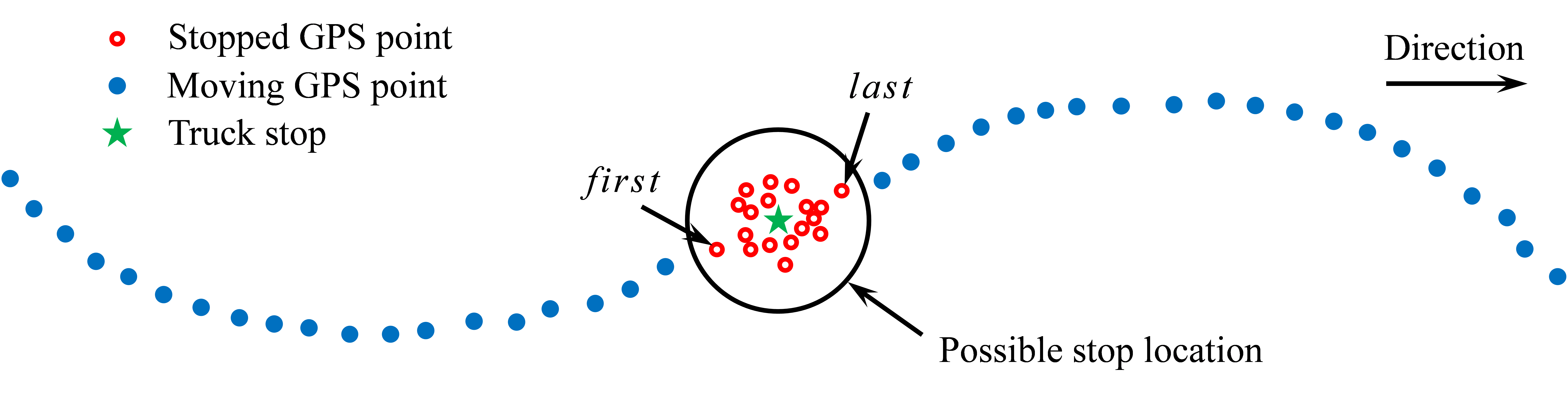}
	\captionsetup{justification=centering}
	\caption{An example of the identification of a truck stop. The black circle marks a possible stop location identified from successive stopped GPS points. The green solid five-pointed star denotes the truck stop, which is located by calculating the average longitude and latitude of the stopped GPS points.}
\end{figure}

To understand the motion characteristics of heavy trucks, we randomly extracted the trajectory data of 100,000 heavy trucks and calculated the speed v between pairs of successive GPS points. The distribution of v is shown in Fig. 3, which shows that the low-speed part of the distribution curve fluctuates greatly without an obvious pattern, while the high-speed part is nearly smooth with a peak on the right side. A stationary truck will also have a small random nonzero speed due to data drift, which is largely responsible for the irregularity of the low-speed part of the distribution. Therefore, we select the intermediate transition point (1.1 km/h) as the speed threshold to eliminate the negative impacts of data drift as much as possible.

\begin{figure}
	\centering
	\includegraphics[scale=0.5]{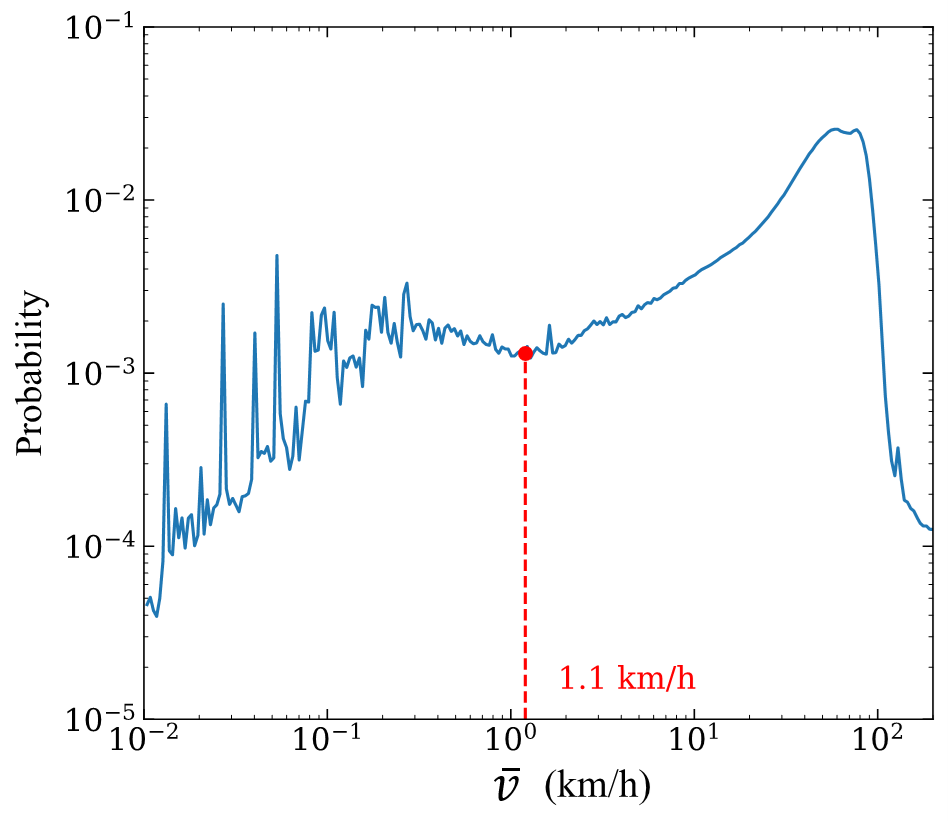}
	\captionsetup{justification=centering}
	\caption{Statistical distribution of the average speed v. The red solid dot denotes the intermediate transition point (1.1 km/h) between the low-speed and high-speed parts of the distribution curve.}
\end{figure}

\subsection{Selecting valid trip ends from among the identified truck stops}
In practice, a heavy truck may stop at certain locations for various reasons (such as loading, unloading, temporary rests or traffic congestion) during a freight trip. We need to select the valid trip ends from among all identified truck stops by inferring the stop purpose of the heavy truck. First, we define minimum and maximum time thresholds by analyzing the distribution of the dwell times of heavy trucks at stop locations and classify truck stops into three types based on these time thresholds. Second, we use highway network GIS data and freight-related POI data to identify valid trip ends from among the three types of truck stops.

\subsubsection{Defining time thresholds}
We use time thresholds to select trip ends from among the identified truck stops. If a heavy truck dwells at a location longer than the defined minimum time threshold, the truck stop is identified as a possible trip end (origin or destination). The dwell times of heavy trucks at stop locations constitute the basic information used to define the time thresholds. We first identified all truck stops using the selected speed threshold of 1.1 km/h and then calculated the dwell times of the heavy trucks at the identified stop locations. The probability distribution of the dwell times is shown in Fig. 4, which shows that the data for short dwell times obey a broken power-law distribution [44], with two power-law segments broken at the point of 24 min. For long dwell times (more than 13 h), the distribution fluctuates greatly with no obvious pattern. Therefore, we define minimum and maximum time thresholds of 24 min and 13 h, respectively, which are the characteristic turning points of the distribution. According to the defined time thresholds, we classify truck stops into three types, i.e., short-term stops (with a dwell time of less than 24 min), medium-term stops (with a dwell time of more than 24 min and less than 13 h) and long-term stops (with a dwell time of more than 13 h).

Next, we need to select valid trip ends from among the three types of truck stops. First, a short-term stop is not a valid trip end because the dwell time of the heavy truck is too short for the completion of one loading or unloading process. Second, a long-term stop is a valid trip end because in such a case, it is likely that the heavy truck is remaining at the current location because the driver's work has ended. Third, medium-term stops may be either valid trip ends or intermediate stops (non-origin/destination stops) caused by a driver’s temporary rest, truck refueling or traffic congestion. Therefore, we need to identify the valid trip ends from among the medium-term stops.

\begin{figure}
	\centering
	\includegraphics[scale=0.5]{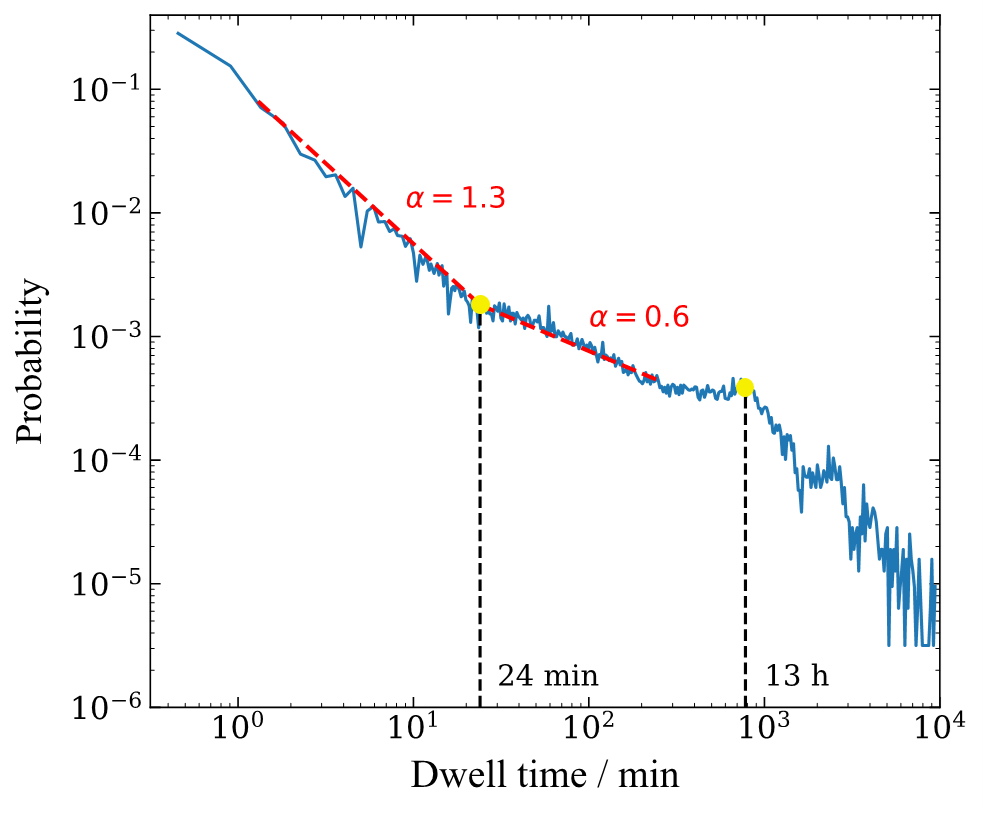}
	\captionsetup{justification=centering}
	\caption{Distribution of heavy truck dwell times at stop locations. The two successive red dashed lines represent two fitted power-law segments with power exponents ${\alpha}$ of 1.3 and 0.6, respectively. The two solid yellow dots denote the characteristic turning points of the distribution. We define minimum and maximum time thresholds of 24 min and 13 h, respectively.}
\end{figure}

\subsubsection{Identifying valid trip ends from among medium-term stops}
We use POI data to determine whether a medium-term stop is located at a freight-related location. Since the POI data contain only the geographic coordinates and semantic description of each location, we first need to detect POI boundaries that are suitable for heavy trucks. For this purpose, we obtained POI data from Amap (https://lbs.amap.com/api/), which is a leading map application in China. In accordance with their relevance to the freight activities of heavy trucks, we selected a variety of freight-related POIs, as shown in Table 2.

\begin{table}[htbp]
	\small
	\centering
	\caption{Details of freight-related POIs}
	\begin{tabular}{cccc}
		\toprule
		POI category              & Number  & Valid radius & POI radius \\
		\midrule
				construction company      & 52,621  & 370 m        & 690 m      \\
		machinery electronics     & 88,336  & 345 m        & 655 m      \\
		chemical metallurgy       & 6,324   & 290 m        & 545 m      \\
		commercial trade          & 86,484  & 275 m        & 516 m      \\
		logistics warehouse       & 4,095   & 260 m        & 487 m      \\
		mining company            & 6,222   & 285 m        & 521 m      \\
		factory                   & 194,476 & 350 m        & 670 m      \\
		farming base              & 91,410  & 310 m        & 550 m      \\
		industrial park           & 67,427  & 450 m        & 849 m      \\
		residential area          & 654,613 & 430 m        & 814 m      \\
		building materials market & 656     & 390 m        & 715 m      \\
		\bottomrule
	\end{tabular}%
\end{table}%

In previous studies on the detection of POI boundaries suitable for pedestrians, a POI boundary has usually been defined in accordance with the building outline. However, for heavy trucks, we must consider more complex circumstances when detecting POI boundaries. A heavy truck may load or unload goods outside the building outline due to the limits of the building space and road conditions. Therefore, we need to comprehensively understand the geospatial relationship between truck stops and nearby POIs. For this purpose, we extracted all truck stops and calculated the distance between each truck stop and the nearest POI for each POI category. Here, we select the POI category “factory” as an example for illustration. The distribution of the distance between a truck stop and the center of the nearest factory is shown in Fig. 5. The probability distribution (see Fig. 5a) has obviously unimodal shape. We define the distance corresponding to the peak probability as the valid radius and define the circle with this valid radius as the valid boundary (see Fig. 6). Most heavy trucks will stop near the valid boundary of a POI. If a truck stop is within this valid boundary, the heavy truck is considered to be stopping at this location. However, a heavy truck may also load or unload goods outside the valid boundary, so we further expand the valid boundary to the POI boundary (the circle with the POI radius). We define the POI radius as the radius where the cumulative probability is twice that at the valid radius in order to calculate the POI radius (see Fig. 5b). Thus, we can obtain a POI boundary for each freight-related POI category.

\begin{figure}
	\centering
	\includegraphics[scale=0.7]{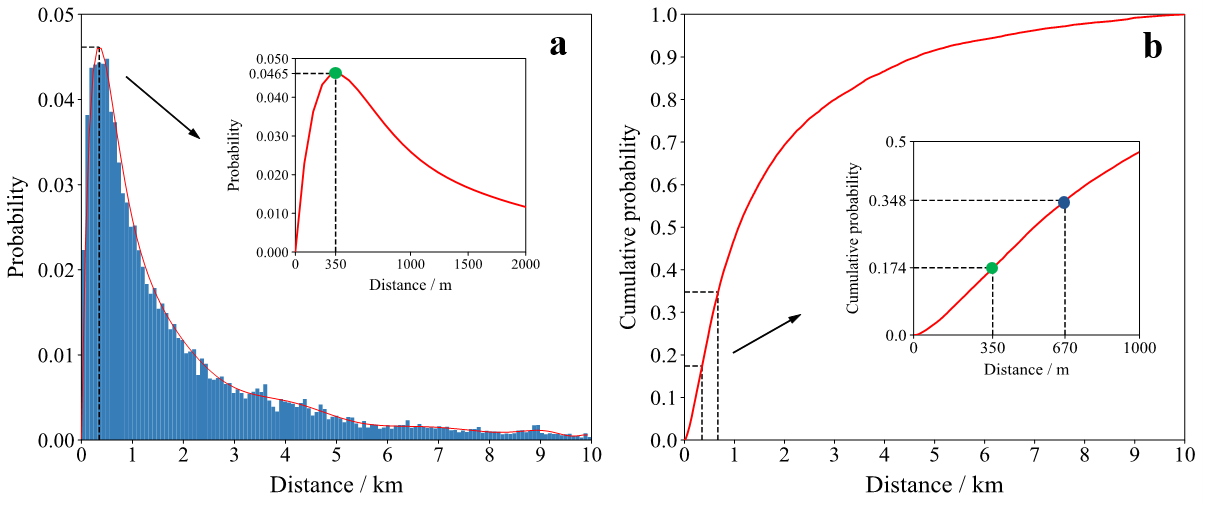}
	\captionsetup{justification=centering}
	\caption{Distribution of the distance between a truck stop and the center of the nearest factory. a. Probability distribution. The subgraph shows the distribution at distances of 0-2 km. The green solid dot indicates the distance (350 m) corresponding to the peak probability. b Cumulative probability distribution. The subgraph shows the distribution at distances of 0-1 km. The green and blue solid dots indicate the valid radius (350 m) and the POI radius (670 m), respectively. The cumulative probability at the POI radius is twice that at the valid radius.}
\end{figure}

\begin{figure}
	\centering
	\includegraphics[scale=1]{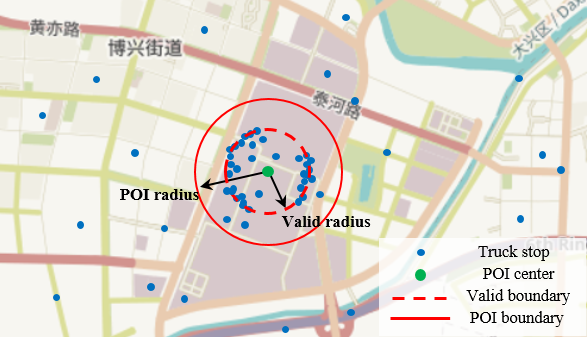}
	\captionsetup{justification=centering}
	\caption{Illustration of the detection of a POI boundary. The central purple area represents a freight-related location. The valid boundary and the POI boundary are circles at the center of the POI with the valid radius and the POI radius, respectively.}
\end{figure}

We extracted all medium-term stops within freight-related POI boundaries, because in such cases, it is likely that the heavy truck is stopping at its current location for loading, unloading or driver’s rest. However, the area within a freight-related POI boundary may also contain urban roads, as shown in Fig. 6. Thus, we need to further filter out the intermediate stops caused by traffic congestion from among these medium-term stops. In this paper, we use highway network GIS data to determine whether the heavy trucks are stuck on the road due to traffic congestion. The highway network GIS data were obtained from OpenStreetMap (www.openstreetmap.org), which is an open-source digital map. We extracted GIS data for four classes of roads (motorways, primary roads, secondary roads and tertiary roads), considering the requirements of heavy trucks in terms of road conditions. According to the design standards for roads in China, both primary and secondary roads have at least four vehicle lanes (where the width of each lane is at least 3.5 m) in one direction, while both motorways and tertiary roads have at least two. In addition, greenbelts, cycleways, footways and emergency lanes may be built on either side of the road. Therefore, we set the average width of primary and secondary roads to 17.5 m, which is equal to a total of five lane widths, and we set the average width of motorways and tertiary roads to 10.5 m, which is equal to a total of three lane widths. A heavy truck is considered stuck on the road due to traffic congestion if the distance between the truck stop and the road centerline is less than half of the average road width.

\subsection{Validation}
The most applicable data for method validation are the travel diaries of heavy trucks, which are difficult to obtain in most cases [12]. Therefore, most studies [14, 20, 41] have adopted land use data and satellite images for method validation. However, land use data can provide only geographic information for polygonal areas, which is not sufficiently detailed for validation. Therefore, in this paper, we use POI data and satellite images to validate the proposed method of freight trip end identification.
Here, we show an example of method validation for one heavy truck in Fig. 7. We can directly determine whether the trip ends are accurately identified using only satellite images if the trip ends are located at freight enterprises with obvious building features (see Fig. 7b). However, when the building features of some freight enterprises are not obvious, using satellite images alone is not sufficient for method validation, and POI data are needed (see Fig. 7c). Moreover, some intermediate stops are incorrectly identified as trip ends, as shown in the example in Fig. 7d; this intermediate stop is located at an expressway rest area inside a freight-related POI boundary, so it is not a valid trip end. The validation results suggest that 94\% of the trip ends are accurately identified in our validation sample of 1,000 heavy trucks.

\begin{figure}
	\centering
	\includegraphics[scale=0.7]{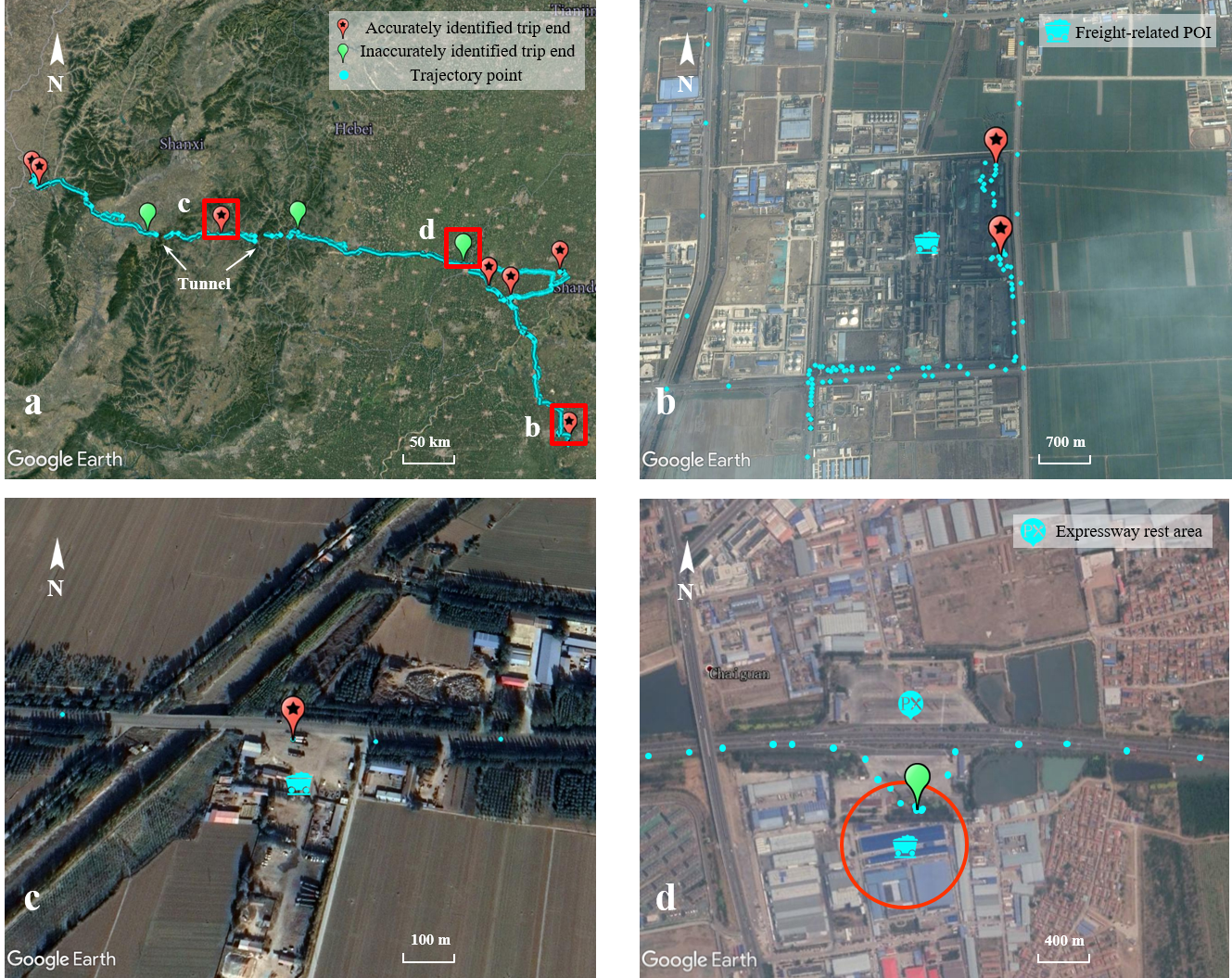}
	\captionsetup{justification=centering}
	\caption{Example of method validation for one heavy truck. a Validation results. The cyan dots represent the trajectory of one heavy truck over one week. The three locations marked in red rectangles correspond to panels b-d. b Only a satellite image is used for method validation. c Both a satellite image and POI data are used for method validation. d Case of an intermediate stop that is incorrectly identified as a trip end. In this case, the heavy truck dwells in an expressway rest area inside a freight-related POI boundary for a long time. The red circle indicates the POI boundary.}
\end{figure}

\section{Results and analysis}
\subsection{Results of intercity freight trip end identification}
We identified 25 million trip ends for 2.6 million heavy trucks using the method introduced above. Then, we extracted the intercity trip ends for each heavy truck based on the administrative divisions of China, which were used to determine whether two successive trip ends were located in two different cities. Finally, we obtained 18 million intercity trip ends from among all trip ends. Furthermore, we analyzed the freight demands of different enterprises by obtaining the POI information of the intercity trip ends, as shown in Table 3. The results show that the freight demands of four types of enterprises, i.e., construction companies, machinery electronics, factories and industrial parks, are obviously larger than the others, implying that heavy trucks play an important role in industrial transportation for these types of enterprises.

\begin{table}[htbp]
	\small
	\centering
	\caption{POI information of intercity freight trip ends}
	\begin{tabular}{cc}
		\toprule
		POI category              & Proportion (\%) \\
		\midrule
		construction company      & 14.65           \\
		machinery electronics     & 12.69           \\
		chemical metallurgy       & 2.57            \\
		commercial trade          & 3.42            \\
		logistics warehouse       & 9.47            \\
		mining company            & 6.03            \\
		factory                   & 22.21           \\
		farming base              & 7.47            \\
		industrial park           & 18.49           \\
		residential area          & 1.31            \\
		building materials market & 1.69             \\
		\bottomrule
	\end{tabular}%
\end{table}%

\subsection{Geographic distribution of intercity freight trips}
We further extracted intercity freight trips of heavy trucks using the identified intercity trip ends. The freight activities of heavy trucks over one week are composed of multiple trip chains, in which two successive trip ends are the origin and destination of the corresponding trip. Thus, we can extract one intercity freight trip by identifying two successive intercity trip ends, as shown in Layer 3 in Fig. 1. We extracted 17 million intercity freight trips, and the geographic distributions of these intercity freight trips at the national and urban agglomeration scales are visualized in Fig. 8. From Fig. 8a, we can see that the freight interaction intensity in the central and eastern areas of mainland China is significantly higher than that in the western area, reflecting regional differences in the distribution of freight demand at the national scale. From Fig. 8b-d, we can see that the freight demand between core cities in an urban agglomeration is distinctly high, while that between other cities is relatively low. The above results help us to understand the economic ties between cities and regions.
\begin{figure}
	\centering
	\includegraphics[scale=0.3]{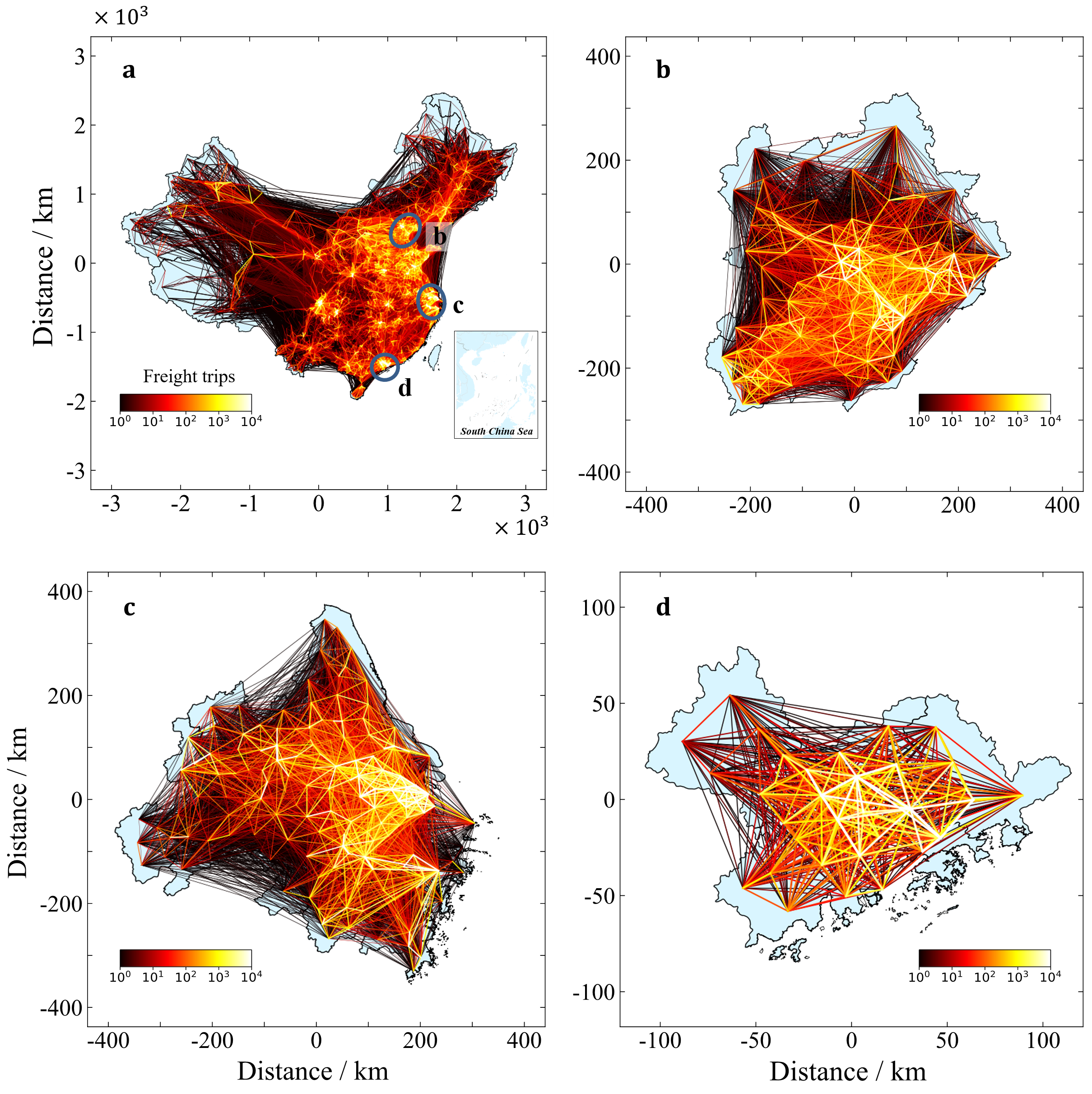}
	\captionsetup{justification=centering}
	\caption{Geographic distribution of intercity freight trips at diverse spatial scales. a Mainland China. Each line represents the freight interaction between two cities. The color of the line indicates the number of freight trips between those two cities. The three areas marked with blue circles correspond to the three urban agglomerations shown in panels b-d. b Beijing-Tianjin-Hebei urban agglomeration. c Yangtze River Delta urban agglomeration. d Pearl River Delta urban agglomeration.}
\end{figure}

\subsection{Spatiotemporal characteristics of intercity freight trips}
The spatiotemporal characteristics of intercity freight trips constitute necessary information for analyzing heavy truck travel behavior [45]. Here, we calculated the trip distance, trip duration, trip departure time and arrival time for each intercity freight trip and obtained the corresponding probability distributions, as shown in Fig. 9. Fig. 9a-b show that the data for both trip distance and trip duration obey lognormal distributions [46], from which we can see that the trip distance of most heavy trucks is approximately 90 km and the trip duration is approximately 3 h. Fig. 9c shows that the freight activities of heavy trucks occur mainly in the daytime, especially between 6:00 and 18:00. The departure time distribution is bimodal (peaks appear at approximately 8:00 and 14:00), while the arrival time distribution is unimodal (a peak appears at approximately 11:00). The above results help us deeply understand the spatiotemporal characteristics of intercity freight trips.
\begin{figure}
	\centering
	\includegraphics[scale=0.4]{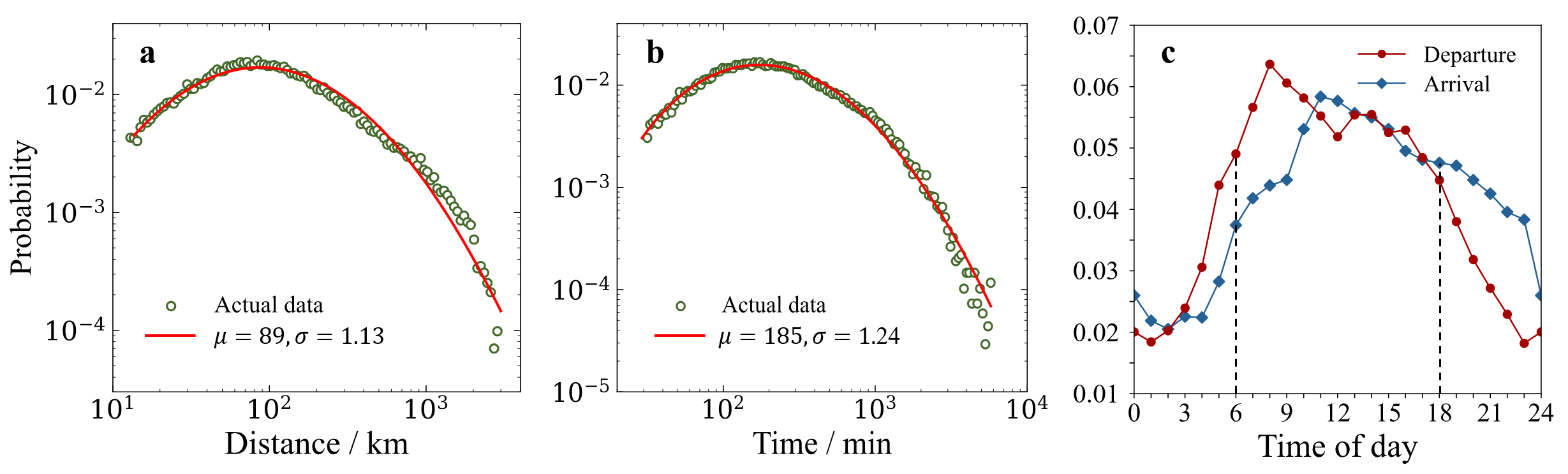}
	\captionsetup{justification=centering}
	\caption{Spatiotemporal characteristics of intercity freight trips. a Distribution of trip distance. The green hollow points represent the actual data, and the red solid line represents the fitted lognormal distribution $1/x/\sigma\sqrt{2\pi}\exp(-(lnx-\mu)^2/2\sigma^2)$. b Distribution of trip duration. c Distributions of trip departure time and arrival time. The two black dashed lines indicate the start and end points of the time period during which freight activities mainly occur.}
\end{figure}

\section{Conclusion and discussion}
In this paper, we propose a data-driven trip end identification method. First, we define a speed threshold of 1.1 km/h by analyzing the speed distribution of heavy trucks. This speed threshold is used to identify truck stops from raw GPS data. Second, we extract all truck stops and obtain the distribution of the heavy truck dwell times at freight-related locations. Accordingly, we define minimum and maximum time thresholds of 24 min and 13 h, respectively, which are the characteristic turning points of the above dwell time distribution. Third, we detect freight-related POI boundaries by analyzing the geospatial relationship between truck stops and nearby POIs. These POI boundaries are used to determine whether a heavy truck is stopping at a certain location. Finally, we use highway network GIS data and freight-related POI data to identify valid trip ends from among the truck stops. In this way, we identify 25 million trip ends from the trajectories of 2.6 million heavy trucks and further extract 18 million intercity trip ends based on the administrative divisions of China. The results of method validation suggest that 94\% of the trip ends are accurately identified. Finally, we analyze the geographic distribution and spatiotemporal characteristics of the 17 million intercity freight trips extracted from the 18 million intercity trip ends.

Intercity freight trip data obtained using our data-driven trip end identification method can be applied in many fields. For example, we can identify the sources of heavy trucks on congested roads using intercity freight trip data to identify the freight bottlenecks of highway networks. These data can help local authorities reduce traffic congestion at bottlenecks and improve traffic efficiency by managing freight demand. Moreover, detailed freight trip information makes it possible to effectively supervise heavy trucks, thus reducing the negative impact of road freight. Moreover, we can gain a deep understanding of the relationship between urban economic development and freight demand by analyzing the geographic distribution of intercity freight trips. We can also use intercity freight trip data to analyze the economic ties between cities [47], thus providing data support for the government to formulate economic policies.

Our data-driven trip end identification method has room for further improvement. For example, although the accuracy of the proposed method is high, if we can obtain heavy truck travel diaries or business orders in the future, we will be able to construct datasets with labels such as delivery, pick-up, and rest; then, we will be able to use machine learning methods [48, 49] to identify the trip ends of heavy trucks to further improve the identification accuracy. Moreover, in this paper, we have used POI data and satellite images to validate our trip end identification method, which is a time-consuming and inevitably subjective process. Access to detailed heavy truck travel diaries could improve the efficiency of validation.

\section{Reference}
[1] L. Yang, X. Lu, Study on the Intercity Highway Freight Network in Beijing-Tianjin-Hebei Region,  2019 16th International Conference on Service Systems and Service Management (ICSSSM), IEEE, 2019, pp. 1-6.

[2] E. Mulholland, J. Teter, P. Cazzola, Z. McDonald, B.P.O. Gallachoir, The long haul towards decarbonising road freight - A global assessment to 2050, Applied Energy, 216 (2018) 678-693.

[3] H. Leridon, World population outlook: Explosion or implosion?, Population \& Societies, 573 (2020) 1-4.

[4] R. Engstrom, The roads' role in the freight transport system, in: L. Rafalski, A. Zofka (Eds.) Transport Research Arena Tra20162016, pp. 1443-1452.

[5] E. Dernir, T. Bektas, G. Laporte, A review of recent research on green road freight transportation, European Journal of Operational Research, 237 (2014) 775-793.

[6] S. Moridpour, E. Mazloumi, M. Mesbah, Impact of heavy vehicles on surrounding traffic characteristics, Journal of Advanced Transportation, 49 (2015) 535-552.

[7] G. Zhang, Y. Li, M.J. King, Q. Zhong, Overloading among crash-involved vehicles in China: identification of factors associated with overloading and crash severity, Injury Prevention, 25 (2019) 36-46.

[8] L. Lyons, A. Lozano, F. Granados, A. Guzman, Impacts of time restriction on heavy truck corridors: The case study of Mexico City, Transportation Research Part a-Policy and Practice, 102 (2017) 119-129.

[9] A.B. Zanjani, A.R. Pinjari, M. Kamali, A. Thakur, J. Short, V. Mysore, S.F. Tabatabaee, Estimation of Statewide Origin-Destination Truck Flows from Large Streams of GPS Data Application for Florida Statewide Model, Transportation Research Record, DOI 10.3141/2494-10(2015) 87-96.

[10] M. Outwater, N. Islam, B. Spear, The magnitude and distribution of commercial vehicles in urban transportation,  84th Transportation Research Board Annual Meeting-Compendium of Papers CDROM, 2005.

[11] S. Hadavi, S. Verlinde, W. Verbeke, C. Macharis, T. Guns, Monitoring Urban-Freight Transport Based on GPS Trajectories of Heavy-Goods Vehicles, Ieee Transactions on Intelligent Transportation Systems, 20 (2019) 3747-3758.

[12] K. Gingerich, H. Maoh, W. Anderson, Classifying the purpose of stopped truck events: An application of entropy to GPS data, Transportation Research Part C-Emerging Technologies, 64 (2016) 17-27.

[13] X. Ma, E.D. McCormack, Y. Wang, Processing Commercial Global Positioning System Data to Develop a Web-Based Truck Performance Measures Program, Transportation Research Record, DOI 10.3141/2246-12(2011) 92-100.

[14] X. Ma, Y. Wang, E. McCormack, Y. Wang, Understanding Freight Trip-Chaining Behavior Using a Spatial Data-Mining Approach with GPS Data, Transportation Research Record, DOI 10.3141/2596-06(2016) 44-54.

[15] H. Oka, Y. Hagino, T. Kenmochi, R. Tani, R. Nishi, K. Endo, D. Fukuda, Predicting travel pattern changes of freight trucks in the Tokyo Metropolitan area based on the latest large-scale urban freight survey and route choice modeling, Transportation Research Part E-Logistics and Transportation Review, 129 (2019) 305-324.

[16] T. Siripirote, A. Sumalee, H.W. Ho, Statistical estimation of freight activity analytics from Global Positioning System data of trucks, Transportation Research Part E-Logistics and Transportation Review, 140 (2020).

[17] B.W. Sharman, M.J. Roorda, Multilevel modelling of commercial vehicle inter-arrival duration using GPS data, Transportation Research Part E-Logistics and Transportation Review, 56 (2013) 94-107.

[18] J. de Vries, R. de Koster, S. Rijsdijk, D. Roy, Determinants of safe and productive truck driving: Empirical evidence from long-haul cargo transport, Transportation Research Part E-Logistics and Transportation Review, 97 (2017) 113-131.

[19] L. Gong, H. Sato, T. Yamamoto, T. Miwa, T. Morikawa, Identification of activity stop locations in GPS trajectories by density-based clustering method combined with support vector machines, Journal of Modern Transportation, 23 (2015) 202-213.

[20] A. Thakur, A.R. Pinjari, A.B. Zanjani, J. Short, V. Mysore, S.F. Tabatabaee, Development of Algorithms to Convert Large Streams of Truck GPS Data into Truck Trips, Transportation Research Record, DOI 10.3141/2529-07(2015) 66-73.

[21] L. Sarti, L. Bravi, F. Sambo, L. Taccari, M. Simoncini, S. Salti, A. Lori, Stop Purpose Classification from GPS Data of Commercial Vehicle Fleets, in: R. Gottumukkala, X. Ning, G. Dong, V. Raghavan, S. Aluru, G. Karypis, L. Miele, X. Wu (Eds.) 2017 17th Ieee International Conference on Data Mining Workshops2017, pp. 280-287.

[22] X. Yang, Z. Sun, X.J. Ban, J. Holguin-Veras, Urban Freight Delivery Stop Identification with GPS Data, Transportation Research Record, DOI 10.3141/2411-07(2014) 55-61.

[23] L.O. Alvares, V. Bogorny, B. Kuijpers, J.A.F.d. Macedo, B. Moelans, A. Vaisman, A model for enriching trajectories with semantic geographical information,  Proceedings of the 15th annual ACM international symposium on Advances in geographic information systems, Association for Computing Machinery, Seattle, Washington, 2007, pp. Article 22.

[24] A.T. Palma, V. Bogorny, B. Kuijpers, L.O. Alvares, A clustering-based approach for discovering interesting places in trajectories,  Proceedings of the 2008 ACM symposium on Applied computing, Association for Computing Machinery, Fortaleza, Ceara, Brazil, 2008, pp. 863–868.

[25] Y. Takimoto, K. Sugiura, Y. Ishikawa, Extraction of Frequent Patterns Based on Users' Interests from Semantic Trajectories with Photographs,  Proceedings of the 21st International Database Engineering \& amp; Applications Symposium, Association for Computing Machinery, Bristol, United Kingdom, 2017, pp. 219–227.

[26] Y. Yang, J. Cai, H. Yang, J. Zhang, X. Zhao, TAD: A trajectory clustering algorithm based on spatial-temporal density analysis, Expert Systems with Applications, 139 (2020).

[27] C. Zhou, D. Frankowski, P. Ludford, S. Shekhar, L. Terveen, Discovering personally meaningful. places: An interactive clustering approach, Acm Transactions on Information Systems, 25 (2007).

[28] X.-l. Zhao, W.-x. Xu, I.C. Soc, A CLUSTERING-BASED APPROACH FOR DISCOVERING INTERESTING PLACES IN A SINGLE TRAJECTORY, 2009.

[29] W. Chen, M. Ji, J. Wang, T-DBSCAN: A Spatiotemporal Density Clustering for GPS Trajectory Segmentation, International Journal of Online Engineering, 10 (2014).

[30] P.F. Laranjeiro, D. Merchan, L.A. Godoy, M. Giannotti, H.T.Y. Yoshizaki, M. Winkenbach, C.B. Cunha, Using GPS data to explore speed patterns and temporal fluctuations in urban logistics: The case of Sao Paulo, Brazil, Journal of Transport Geography, 76 (2019) 114-129.

[31] P. Camargo, S. Hong, V. Livshits, Expanding the Uses of Truck GPS Data in Freight Modeling and Planning Activities, Transportation Research Record, DOI 10.3141/2646-08(2017) 68-76.

[32] Z. Cheng, L. Jiang, D. Liu, Z. Zheng, Ieee, Density based spatio-temporal trajectory clustering algorithm,  Igarss 2018 - 2018 Ieee International Geoscience and Remote Sensing Symposium2018, pp. 3358-3361.

[33] S. Louhichi, M. Gzara, H. Ben Abdallah, Ieee, A density based algorithm for discovering clusters with varied density, 2014.

[34] T. Luo, X. Zheng, G. Xu, K. Fu, W. Ren, An Improved DBSCAN Algorithm to Detect Stops in Individual Trajectories, Isprs International Journal of Geo-Information, 6 (2017).

[35] D. Ashbrook, T. Starner, Using GPS to learn significant locations and predict movement across multiple users, Personal and Ubiquitous Computing, 7 (2003) 275-286.

[36] L. Gong, T. Yamamoto, T. Morikawa, Identification of activity stop locations in GPS trajectories by DBSCAN-TE method combined with support vector machines, in: P. Bonnel, M. Munizaga, C. Morency, M. Trepanier (Eds.) Transport Survey Methods in the Era of Big Data: Facing the Challenges2018, pp. 146-154.

[37] Z. Fu, Z. Tian, Y. Xu, C. Qiao, A Two-Step Clustering Approach to Extract Locations from Individual GPS Trajectory Data, Isprs International Journal of Geo-Information, 5 (2016).

[38] S.P. Greaves, M.A. Figliozzi, Collecting Commercial Vehicle Tour Data with Passive Global Positioning System Technology:Issues and Potential Applications, Transportation Research Record, 2049 (2008) 158-166.

[39] E. McCormack, M.E. Hallenbeck, ITS Devices Used to Collect Truck Data for Performance Benchmarks, Transportation Research Record, 1957 (2006) 43-50.

[40] R. Friswell, A. Williamson, Management of heavy truck driver queuing and waiting for loading and unloading at road transport customers' depots, Safety Science, 120 (2019) 194-205.

[41] R. Aziz, M. Kedia, S. Dan, S. Basu, S. Sarkar, S. Mitra, P. Mitra, Identifying and Characterizing Truck Stops from GPS Data, in: P. Perner (Ed.) Advances in Data Mining: Applications and Theoretical Aspects2016, pp. 168-182.

[42] S. Hess, M. Quddus, N. Rieser-Schuessler, A. Daly, Developing advanced route choice models for heavy goods vehicles using GPS data, Transportation Research Part E-Logistics and Transportation Review, 77 (2015) 29-44.

[43] J. Jun, R. Guensler, J.H. Ogle, Smoothing Methods to Minimize Impact of Global Positioning System Random Error on Travel Distance, Speed, and Acceleration Profile Estimates, Transportation Research Record, 1972 (2006) 141-150.

[44] R. Tomaschitz, Multiply broken power-law densities as survival functions: An alternative to Pareto and lognormal fits, Physica a-Statistical Mechanics and Its Applications, 541 (2020).

[45] A. Amer, J.Y.J. Chow, A downtown on-street parking model with urban truck delivery behavior, Transportation Research Part a-Policy and Practice, 102 (2017) 51-67.

[46] C.C. Heyde, On a property of the lognormal distribution, Journal of the Royal Statistical Society: Series B (Methodological), 25 (1963) 392-393.

[47] Q. Shi, X. Yan, B. Jia, Z. Gao, Freight Data-Driven Research on Evaluation Indexes for Urban Agglomeration Development Degree, Sustainability, 12 (2020).

[48] S. Dabiri, K. Heaslip, Inferring transportation modes from GPS trajectories using a convolutional neural network, Transportation Research Part C-Emerging Technologies, 86 (2018) 360-371.

[49] R. Zhang, P. Xie, C. Wang, G. Liu, S. Wan, Classifying transportation mode and speed from trajectory data via deep multi-scale learning, Computer Networks, 162 (2019).

\end{document}